\documentclass[conference]{IEEEtran}
\IEEEoverridecommandlockouts
\usepackage{cite}
\usepackage{amsmath,amssymb,amsfonts}
\usepackage{algorithmic}
\usepackage{graphicx}
\usepackage{textcomp}
\usepackage{xcolor}
\usepackage[linesnumbered,ruled,vlined]{algorithm2e} %added, for our algorithm

\def\BibTeX{{\rm B\kern-.05em{\sc i\kern-.025em b}\kern-.08em
    T\kern-.1667em\lower.7ex\hbox{E}\kern-.125emX}}

\newcommand{\linebreakand}{%
  \end{@IEEEauthorhalign}
  \hfill\mbox{}\par
  \mbox{}\hfill\begin{@IEEEauthorhalign}
}

\begin{document}

\title{Distance-Only Task Orchestration Algorithm for Energy Efficiency in Satellite-Based Mist Computing}

\author{
\IEEEauthorblockN{
Messaoud Babaghayou \IEEEauthorrefmark{1},
Noureddine Chaib \IEEEauthorrefmark{1},
Leandros Maglaras \IEEEauthorrefmark{2}, \\
Yagmur Yigit \IEEEauthorrefmark{2}, and 
Mohamed Amine Ferrag \IEEEauthorrefmark{3}}
\IEEEauthorblockA{
\IEEEauthorrefmark{1} LIM Laboratory. Laghouat University, Laghouat, Algeria \\
\IEEEauthorrefmark{2} School of Computing, Engineering and The Build Environment, Edinburgh Napier University, United Kingdom \\
 \IEEEauthorrefmark{3} Technology Innovation Institute (TII), Abu Dhabi, United Arab Emirates
 \\
Email: messaoud.babaghayou@lagh-univ.dz, n.chaib@lagh-univ.dz, l.maglaras@napier.ac.uk, \\ yagmur.yigit@napier.ac.uk, mohamed.ferrag@tii.ae }

}

\maketitle

\begin{abstract}
%This paper presents a task orchestration algorithm designed to enhance energy efficiency in satellite-based mist computing systems. The proposed algorithm leverages distance as the primary criterion for task offloading, with a focus on optimizing resource allocation in the three-tier architecture: cloud satellites, edge datacenters, and mist nodes. Compared versus four other algorithms, the distance-based approach is shown to significantly reduce energy consumption, making it an ideal solution for resource-constrained satellite environments.
This paper addresses the challenge of efficiently offloading heavy computing tasks from ground mobile devices to the satellite-based mist computing environment. With ground-based edge and cloud servers often being inaccessible, the exploitation of satellite mist computing becomes imperative. 
Existing offloading algorithms have shown limitations in adapting to the unique characteristics of heavy computing tasks. Thus, we propose a heavy computing task offloading algorithm that prioritizes satellite proximity. This approach not only reduces energy consumption during telecommunications but also ensures tasks are executed within the specified timing constraints, which are typically non-time-critical.
Our proposed algorithm outperforms other offloading schemes in terms of satellites energy consumption, average end-to-end delay, and tasks success rates. Although it exhibits a higher average VM CPU usage, this increase does not pose critical challenges. This distance-based approach offers a promising solution to enhance energy efficiency in satellite-based mist computing, making it well-suited for heavy computing tasks demands.
\end{abstract}


\section{Introduction}
\label{s1}

In recent years, we have witnessed a profound transformation in the landscape of mobile computing, catalyzed by the convergence of two compelling visions: the Internet of Things (IoT) and the advent of 5G communications. This paradigm shift has steered mobile computing away from the traditional centralized Mobile Cloud Computing model towards the realms of Mobile Edge Computing (MEC)~\cite{mao2017survey,babaghayou2023safety}.
From another corner, over four decades ago, the pioneers of satellite communication, J. R. Pierce and R. Kompfner, embarked on a visionary exploration of transoceanic communication possibilities through satellite technology. Their seminal work delved into myriad considerations, encompassing alternative satellite repeater concepts, intricate discussions on orbital dynamics, the optimization of satellite constellations, the intricacies of path-loss calculations, the evolution of modulation systems, and the persistent unknowns in satellite communications. It is remarkable that despite the passage of almost 40 years, the core principles they laid down have been the foundation for the field~\cite{wu1997satellite}.

\begin{figure}[t]
    \centering
    \includegraphics[width=3in]{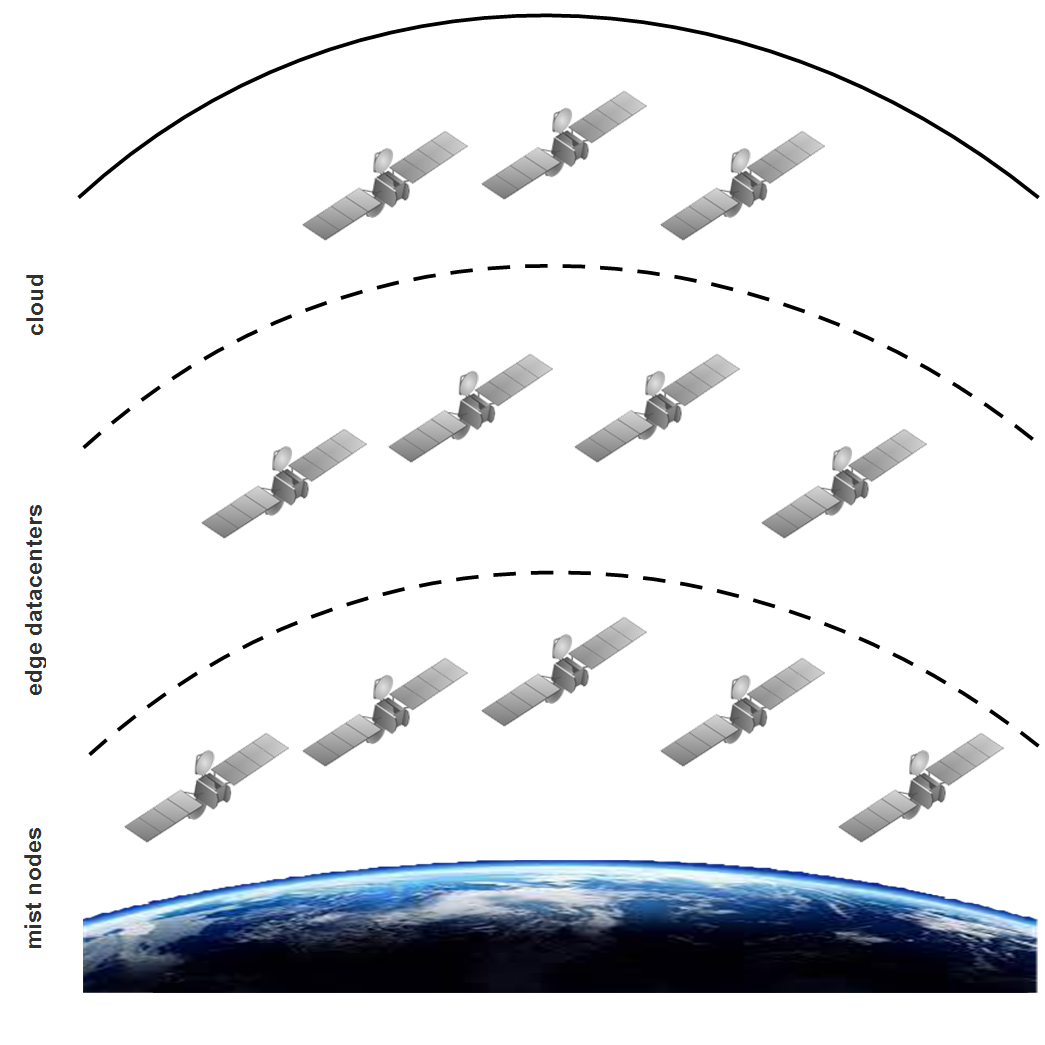}
    \caption{The general satellite edge computing architecture.}
    \label{fig:st}
\end{figure}

The ascent of IoT and satellite-to-satellite communication has become an undeniable reality, charting a promising trajectory for the satellite market. Presenting a transformative narrative, a comprehensive report from global industry analysts underscores the significance of satellite-to-satellite IoT services. This paradigm shift is fueled by the ever-expanding needs of global enterprises and governments, who seek to monitor, manage, and assert control over geographically dispersed assets, both stationary and mobile. The satellite-to-satellite IoT services market is poised for substantial growth. Evidently, satellite operators are actively responding to this burgeoning demand by offering comprehensive IoT solution packages, harnessing the capabilities of satellite technology~\cite{de2015satellite}.
As a consequence, edge computing within the satellite domain has garnered increasing attention from various stakeholders. This approach comprises distinct layers, as shown in Fig.~\ref{fig:st}, including the mist or edge devices layer, followed by the data centers layer, culminating in the cloud layer, each representing progressively higher altitudes.

%%%%%
%%%%%

%The concept of ``heavy computing tasks'' has become prevalent, demanding substantial computational power and posing challenges for devices with limited resources. In UAVs, these tasks involve video preprocessing, pattern recognition, and feature extraction, requiring complex algorithms and powerful processors \cite{messous2019game}. Similarly, in the integration of blockchain with IoT, heavy computing arises during the mining process, placing significant demand on storage and computing capacities \cite{yan2020edge}. These challenges emphasise the need to offload tasks to more potent servers or nearby devices for efficient processing. We propose a novel approach for offloading heavy computing tasks to satellite-based mist computing, prioritizing proximity to satellites to address these challenges. This work offers a unique solution that enhances energy efficiency and performance for heavy computing tasks.

Within this transformative landscape of mobile and satellite communications, the demand for "heavy computing tasks" emerges as a crucial consideration. These tasks encompass a spectrum of compute-intensive operations, ranging from intricate calculations involved in satellite communication optimizations to the demanding processing needs of blockchain-based IoT frameworks. In UAVs, these tasks involve video preprocessing, pattern recognition, and feature extraction, requiring substantial computational capabilities \cite{messous2019game}. On the other hand, in blockchain integration within IoT, heavy computing arises during the mining process, demanding significant processing power \cite{yan2020edge}. The authors' research delves into the efficient handling of such computationally intensive tasks within the satellite-based mist computing environment, aiming to optimize energy consumption and ensure timely task execution. This novel approach represents a pivotal step towards accommodating heavy computing tasks and fostering a more efficient satellite-based mist computing paradigm.

It is imperative to emphasize that the meticulous preservation of privacy and security is paramount in IT systems in general \cite{babaghayou2021between, babaghayou2020impact}, and in the realm of satellite-based mist computing in particular \cite{baselt2022security}. These critical aspects serve as the bedrock upon which the success and reliability of the entire system stand. Failing to uphold robust privacy and security measures would render the system vulnerable to breaches by adversaries and malicious actors, potentially leading to severe consequences that can undermine the efficacy and trustworthiness of the entire infrastructure. Therefore, a vigilant and proactive approach to safeguarding privacy and security is essential to fortify the foundations of satellite-based mist computing and ensure its continued success. 

Furthermore, the meticulous preservation of Quality of Service (QoS), encompassing crucial factors such as latency and various types of delays, is vital. This applies not only to general IT systems \cite{dagli2020resiliency} but also holds particular significance within the context of satellite systems \cite{de2015satellite}. The seamless operation and performance of these systems hinge upon the consistent delivery of QoS standards, making it imperative to maintain and optimize these aspects for their sustained effectiveness.

This paper contributes to the field of satellite-based mist computing with the following key contributions:

\begin{itemize}
    \item Distance-Based Task Offloading Algorithm: We introduce a novel algorithm for task orchestration that prioritizes proximity to satellites, allowing for efficient and energy-conscious offloading of heavy computing tasks from ground mobile devices. This approach is tailored to meet the timing requirements of such tasks, which are often non-time-critical.
    
    \item Enhanced Energy Efficiency: Our proposed algorithm significantly reduces energy consumption during task execution by optimizing the use of nearby satellites, making it an environmentally and economically sustainable solution.
    
    \item Improved Tasks Success Rate: Through proximity-based offloading, our algorithm enhances the overall success rate of task execution, ensuring that heavy computing tasks are reliably completed in a satellite-based environment.
    
    \item Reduced End-to-End Delay: We demonstrate that the distance-based approach leads to a substantial reduction in average end-to-end task execution delay, which is critical for meeting performance expectations.    
\end{itemize}

The remainder of this paper is organized as follows:
In Section~\ref{s2}, we review related works in satellite communications technology, establishing the context for our study. Section~\ref{s3} introduces our system model, outlining key components and parameters governing satellite edge computing. Section~\ref{s4} delves into task deployment strategies, including common methods and our distance-only approach, facilitating detailed comparisons. Section~\ref{s5} presents outcomes from extensive experimentation and simulations, analyzing performance metrics such as latency, energy consumption, CPU utilisation and overall tasks success rates. Finally, in Section~\ref{s6}, we summarise our findings and highlight the significance of our proposed approach in optimizing heavy computing tasks for satellite edge computing environments, paving the way for more efficient satellite-based applications and giving future works and directions.

\section{Related Work}
\label{s2}

A recent study by Wang et al.~\cite{wang2021dynamic} explores the potential of Low Earth Orbit (LEO) satellites in satellite-based edge computing. LEO satellites offer wide coverage and low latency, making them valuable for providing computing services to user access terminals. The study focuses on resource allocation challenges in Edge Computing Satellites (ECS) due to diverse resource requirements and access planes. It proposes a three-layer network architecture with a software-defined networking (SDN) model to manage inter-satellite links (ISLs) and ECS resource scheduling. Advanced algorithms like the Advanced K-means Algorithm (AKG) and a breadth-first-search-based spanning tree algorithm (BFST) are employed for resource allocation and ISL construction. Simulation results affirm the feasibility and effectiveness of this dynamic resource scheduling scheme, addressing key challenges in satellite-based edge computing systems.

Zhang et al. conducted a study to explore the vital integration of satellite and terrestrial networks in the context of 6G wireless architectures~\cite{zhang2020double}. These integrated networks are essential for delivering robust and secure connectivity over expansive geographic areas. The authors introduce the concept of double-edge intelligent integrated satellite and terrestrial networks (DILIGENT), emphasizing the need for synergizing communication, storage, and computation capabilities between satellite and cellular networks. Leveraging multi-access edge computing (MEC) technology and artificial intelligence (AI), the DILIGENT framework is designed for systematic learning and adaptive network management. The article provides an overview of academic research and standardization efforts, followed by an in-depth exploration of the DILIGENT architecture, highlighting its advantages. Strategies such as task offloading, content caching, and distribution are discussed, with numerical results demonstrating the superior performance of this network architecture compared to existing integrated networks.

The authors in~\cite{wang2019satellite} delve into the growing significance of the Internet of Things (IoT) in the information industry. To address challenges related to network distance and remote IoT device deployment, edge computing emerges as a promising paradigm. In scenarios where IoT devices are situated in remote or inaccessible areas, relying on satellite communication becomes imperative. However, conventional satellites are often specialized and lack universal computing capabilities. The proposed solution involves transforming traditional satellites into space edge computing nodes, enabling dynamic software loading, resource sharing, and coordinated services with the cloud. The article outlines the hardware structure and software architecture of such satellites and presents modelling and simulation results. Findings indicate that the space edge computing system exhibits superior performance in terms of reduced time and energy consumption compared to traditional satellite constellations, with service quality influenced by satellite quantity, performance, and task offloading strategies.

Gao et al.~\cite{gao2022virtual} focus on the role of satellite networks as complements to terrestrial networks, catering to the computing needs of Internet of Things (IoT) users in remote regions. Given the inherent limitations of satellites, including constraints in computing power, storage, and energy, an innovative approach involves breaking down IoT user computation tasks into segments and leveraging multiple satellites for collaborative processing to enhance satellite network efficiency. The integration of Network Function Virtualization (NFV) technology with satellite edge computing is an emerging area of interest. The paper introduces a potential game-based solution for Virtual Network Function (VNF) placement within satellite edge computing. The objective is to minimize deployment costs for individual user requests while maximizing the provision of computing services across the satellite network. This optimization problem is formulated as a potential game, aiming to maximize overall network benefits through a game-theoretical approach. The proposed decentralized resource allocation algorithm, based on a potential game (PGRA), seeks a Nash equilibrium to effectively address the VNF placement challenge. Experimental simulations validate the efficacy of the PGRA algorithm in solving the VNF placement problem within satellite edge computing.

In their study \cite{babaghayou2023safety}, introduce the OVR scheme, emphasizing its significance in addressing location privacy and road congestion challenges within the Internet of Vehicles (IoV). OVR innovatively leverages silent periods to enhance location privacy and manage congestion in real time. Comparative evaluations against existing privacy schemes highlight OVR's superior performance in privacy and QoS. 

%%This work aligns with the exploration of efficient computing task offloading in satellite-based mist computing, as both endeavours seek to optimize resource allocation and enhance these evolving paradigms.

\section{System Model}
\label{s3}

In order to provide a comprehensive depiction of the system model employed, we elucidate the fundamental components taken into account both during the formulation of our approach and its subsequent evaluation. These components are stated as follows:

\begin{itemize}
\item Generated heavy computing tasks:  They are the tasks that are being generated from ground-level sources like IoT devices, UAVs \cite{yigit2023twinport} or mobile devices and are typically processed at the satellite level. We assume that it starts from the mist nodes level. Our work addresses the challenge of efficiently managing these tasks by prioritizing proximity to satellites in task offloading. This ensures that heavy computing tasks are processed efficiently while considering their origin and timing requirements.
\item Mist nodes: they refer to the integral components of the satellite-based mist computing infrastructure. These nodes serve as the intermediary layer, responsible for executing heavy computing tasks offloaded from ground devices or relaying them to cloud satellites or edge datacenters. They play a pivotal role in optimizing the processing of tasks, reducing latency, and enhancing the overall efficiency of the mist computing system.
\item Edge datacenters are essential components, serving as critical hubs for processing data and heavy computing tasks. These datacenters play a key role in our three-tier architecture, offering a balance between local processing and centralized cloud resources. They help reduce latency, enhance responsiveness, and facilitate efficient task execution in satellite-based mist computing, ultimately contributing to improved system performance. They also may run Artificial Intelligence (AI) techniques to achieve various objectives \cite{babaghayou2023ai}
\item Cloud computing centers are central facilities in the satellite-based mist computing architecture. They provide the vast computational resources needed to support heavy computing tasks and data processing. While they may not be as geographically proximate as edge datacenters, they serve as essential backbones for the system, ensuring scalability and reliability. These centers are a crucial component in our three-tier architecture, contributing to the overall efficiency and effectiveness of satellite-based mist computing.
\end{itemize}

\section{Methodology}
\label{s4}

Our methodology is designed to address the unique challenges posed by heavy computing task offloading in a satellite-based mist computing environment. We emphasize the need to efficiently manage tasks originating from ground-level sources, including IoT devices, UAVs, and other mobile devices. The primary objective of our methodology is to optimize energy efficiency, reduce end-to-end delay, and improve tasks success rates in this context.

\subsection{Task Orchestration Algorithm Development}
To achieve efficient task orchestration, we have proposed a novel distance-based algorithm that takes into consideration the closest destination in which the generated heavy computing tasks would be executed. This algorithm focuses on the following two aspects:

\subsubsection{Proximity-Based Offloading}
Our algorithm prioritizes satellites that are geographically closer to the source of the tasks. This ensures that task offloading is energy-efficient, taking into account the practicality of processing tasks generated from ground-level devices. We consider that the tasks are generated from the mist nodes level for simplicity.

\subsubsection{Timing Requirements}
Recognizing that heavy computing tasks are often non-time-critical, we designed the algorithm to handle them within their specified timing constraints, which are large to some extent. Nevertheless, we tend towards lower end-to-end delays for optimum results.

\subsection{Experimental Evaluation}
To assess the effectiveness of our methodology, we conducted a comprehensive experimental evaluation. We compared our distance-based algorithm with existing offloading schemes, considering the following key performance metrics:

\subsubsection{VM CPU Usage}
We monitored the average VM CPU usage, ensuring it remained within acceptable limits and did not introduce critical resource constraints.
\subsubsection{Average End-to-End Delay}
We analyzed the average time it takes for tasks to be successfully completed, taking into account the specific timing requirements of heavy computing tasks that are, in this case scenario, large.
\subsubsection{Satellites Energy Consumption}
We measured the energy consumption of satellites during task execution to evaluate the energy efficiency of our approach.
\subsubsection{Tasks Success Rate}
We assessed the overall success rate of task execution to gauge the reliability of our methodology since this metric is so critical and having a lot of failing tasks would impact the effectiveness of the approach.

\subsection{Task Orchestration Algorithms}
To comprehensively assess the effectiveness of our proposed distance-based algorithm, we have selected four additional task orchestration algorithms that are implemented and used in the work done by Wei et al. in \cite{wei2020satedgesim}. These chosen algorithms serve as benchmarks and are essential for the comparative evaluation alongside our proposed algorithm. They are stated as follows:

\subsubsection{Distance\_Only}
Our Distance\_Only Algorithm, at its core, is designed to optimize task orchestration in a satellite-based mist computing environment. It accomplishes this by prioritizing proximity as the primary factor in deciding where and how to offload heavy computing tasks. By considering the physical distance between the source satellite and potential offloading destinations, this algorithm aims to minimize energy consumption, reduce end-to-end delays and ensure high tasks success rates. This approach ensures that tasks are efficiently processed, especially those originating from ground-level devices, such as IoT sensors, UAVs and mobile devices where their available resources are so restricted. Algorithm \ref{algo:doa} explains the operational principles of our proposed algorithm.

\begin{algorithm}[htbp]
\caption{Distance\_Only Algorithm}
\label{algo:doa}
\KwData{architecture (list of available resources), task (the task to be offloaded)}
\KwResult{Selected VM for Task Offloading}

\BlankLine
Create an empty list 'distance'\;
\ForEach{VM in 'orchestrationHistory'}{
    Calculate the distance delay ('distance\_tem') for the VM and the task\;
    Add 'distance\_tem' to the 'distance' list\;
}
Standardize the 'distance' values and store them in 'distance\_stand'\;
Initialize 'vm' to -1 and 'min' to -1\;
\ForEach{VM in 'orchestrationHistory'}{
    \If{offloading the task to the current VM is possible}{
        Calculate 'min\_factor' based on the standardized delay from 'distance\_stand'\;
        \If{'min' is -1}{
            Set 'min' to 'min\_factor' and 'vm' to the current VM\;
        }
        \ElseIf{'min\_factor' is less than 'min'}{
            Update 'min' to 'min\_factor' and 'vm' to the current VM\;
        }
    }
}
Return the selected VM ('vm') for task offloading\;
\end{algorithm}

\subsubsection{Round\_Robin}
%%The Round\_Robin task deployment scheme
it operates by choosing the satellite node with the fewest assigned tasks, thus promoting a balanced distribution of the computing workload. By prioritizing underutilized resources, it aims to prevent resource bottlenecks and alleviate the strain on heavily loaded nodes. This equitable distribution strategy contributes to optimal system performance and task completion rates. Furthermore, it minimizes the likelihood of overburdening any single satellite node, ensuring that all resources are used efficiently and effectively. In scenarios where task requirements vary, the scheme serves as a fair and pragmatic approach for optimizing task deployment in satellite-based mist computing environments.

\subsubsection{Trade\_Off}
%%The Trade\_Off task deployment scheme
it follows a dynamic approach that intelligently selects the most suitable satellite node based on a nuanced balance of factors, including latency, energy consumption, and resource availability. This algorithm evaluates each candidate satellite node, considering the node's type (cloud or edge device), the number of waiting tasks, available CPU processing power, and specific task characteristics. By calculating a combined weighted score for each candidate, this scheme ensures tasks are allocated to the satellite, offering the most optimal trade-off between these factors. This approach enhances the efficiency and effectiveness of task orchestration, optimizing satellite resource utilization in satellite-based mist computing environments.

\subsubsection{Random\_VM}
%%The Random\_VM task deployment scheme
itintroduces an element of randomness into the selection of the most suitable satellite node for task deployment. It compiles a list of available VMs and utilizes a randomization strategy to assign tasks, promoting load balancing and diversification among satellite nodes. The process randomly selects a VM index and verifies its suitability for task offloading. This approach adds an element of unpredictability to task assignment, reducing the risk of overloading specific nodes. If a suitable VM is not found through random selection, the scheme systematically scans for an eligible VM, ensuring equitable resource utilization and offering a dynamic approach to task deployment in satellite-based mist computing environments.

\subsubsection{WEIGHT\_GREEDY (WG)}
it is a sophisticated approach designed for satellite-based mist computing environments. It incorporates a comprehensive evaluation of four critical performance indicators: transmission distance, CPU processing time, number of parallel tasks, and equipment energy consumption. What sets "WEIGHT\_GREEDY" as a good scheme is its fine-tuned weighting of these indicators, where the ratios are adjusted by its authors to be 6:6:5:3. This means that factors related to task timeliness and minimal energy consumption are prioritized due to their significant weight. In a satellite edge computing scenario, where timely task execution and energy efficiency are paramount, this weighting ratio ensures that the scheme optimally balances the competing demands of low latency and minimal power consumption.

\section{Results and Discussion}
\label{s5}

\subsection{Simulation Environment}
In our simulations, we leveraged the "SatEdgeSim" simulator \cite{wei2020satedgesim}, an innovative toolkit tailored for modeling and simulating performance evaluation in satellite edge computing environments. "SatEdgeSim" is built upon the foundation of the well-established "PureEdgeSim" simulator \cite{mechalikh2021pureedgesim}, which, in turn, relies on the widely recognized "CloudSim Plus" cloud computing simulator \cite{silva2017cloudsim}. This strategic choice streamlines the development process, as it eliminates the need to construct the underlying code for traditional cloud computing and fog, edge or mist computing environments. Instead, "SatEdgeSim" can focus its efforts on creating a specialized satellite edge computing environment that aligns with the functional and performance requirements of this unique domain. This framework enables researchers to conduct comprehensive simulations and evaluate task deployment strategies within a satellite-based mist computing context.

\subsection{Experimental Setup}
A comprehensive overview of the experimental setup is encapsulated within Table \ref{tab:params} for a convenient reference.

\begin{table}[htbp]
\caption{Experimental Setup Parameters}
\label{tab:params}
\centering
\begin{tabular}{|p{4cm}|p{4cm}|}
\hline
\textbf{Parameter} & \textbf{Value} \\
\hline
Simulation Time & 10 minutes \\
Updating Interval & 1 second \\
Satellite Minimum Height & 400,000 meters \\
Maximum Mist Satellites Communication Range & 32,000,000 meters \\
Maximum Edge Datacenter Satellites Communication Range & 36,000,000 meters \\
Maximum Cloud Satellites Communication Range & 40,000,000 meters \\
Number of Mist Satellites & 1000 \\
Number of Edge Datacenter Satellites & 24 \\
Number of Cloud Satellites & 18 \\
Progression by Each Iteration & 100 satellites \\
Tasks Generation Rate & 20 tasks per minute \\
Type of Tasks & Heavy Computing Tasks \\
WLAN Bandwidth & 1000 megabits per second \\
WAN Propagation Speed & 300,000,000 meters per second \\
Coordinate Positions & Determined using the STK tool (Satellite Tool Kit) \\
Consumed Energy & 0.00000005 J per bit \\
Amplifier Dissipation Free Space & 0.00000000001 J per bit per m\textsuperscript{2} \\
Amplifier Dissipation Multipath & 0.0000000000000013 J per bit per m\textsuperscript{4} \\
Orchestration Architectures & Utilize all layers (Mist, Edge Datacenters, Cloud) \\
Orchestration Algorithms & Round\_Robin, Trade\_Off, Random\_VM, WG, Distance\_Only \\
\hline
\end{tabular}
\end{table}

\subsection{Simulation Results}

In this section, we present the results of our extensive evaluation and comparative analysis of five task orchestration algorithms. Specifically, we examined the performance of Round\_Robin, Trade\_Off Random\_VM, WG and our proposed Distance\_Only algorithm. These algorithms were rigorously assessed using a set of critical metrics, including VM CPU Usage, Average End-to-End Delay, Satellites Energy Consumption, and Tasks Success Rate. Our findings shed light on the effectiveness and efficiency of these task orchestration strategies in the context of satellite-based mist computing environments.

\subsubsection{VM CPU Usage}

\begin{figure}
\includegraphics[width=\linewidth]{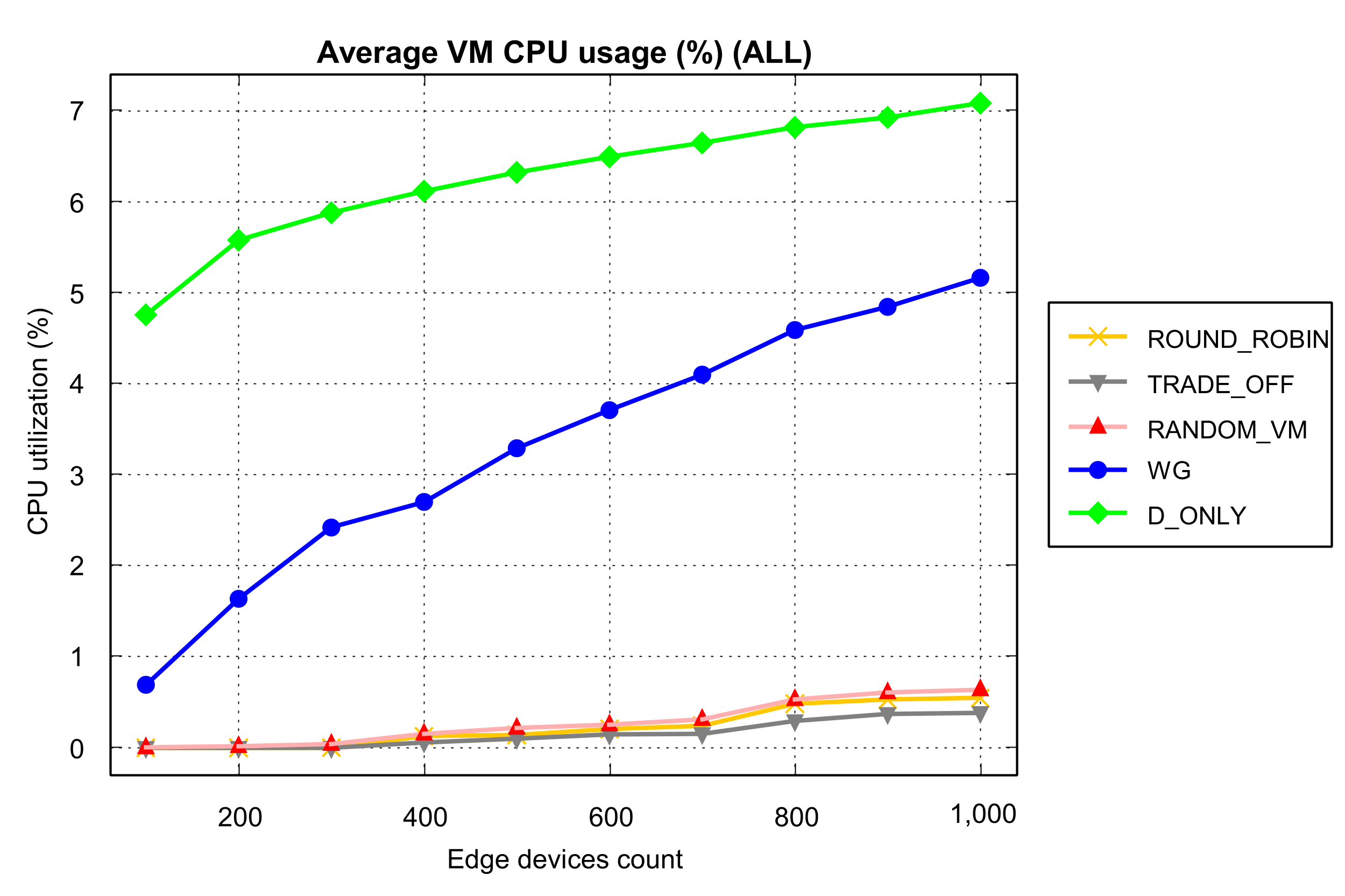}
\caption{VM CPU Usage of the orchestration algorithms}
\label{fig:s1}
\end{figure}

The examination of VM CPU Usage across varying numbers of satellites, as shown in Fig. \ref{fig:s1}, revealed distinctive patterns among the task orchestration algorithms. Notably, Round\_Robin, Trade\_Off and Random\_VM exhibited consistently low CPU utilization, which proportionally increased with the satellite count. In contrast, the WG algorithm consumed a higher CPU utilization, ranging from 0.7 percent to 5.1 percent as the satellite numbers increased. Our proposed Distance\_Only algorithm displayed CPU utilization that initiated at approximately 4.8 percent and gradually rose to 7.1 percent with the expansion of satellite nodes.

\subsubsection{Average End-to-End Delay}

\begin{figure}
\includegraphics[width=\linewidth]{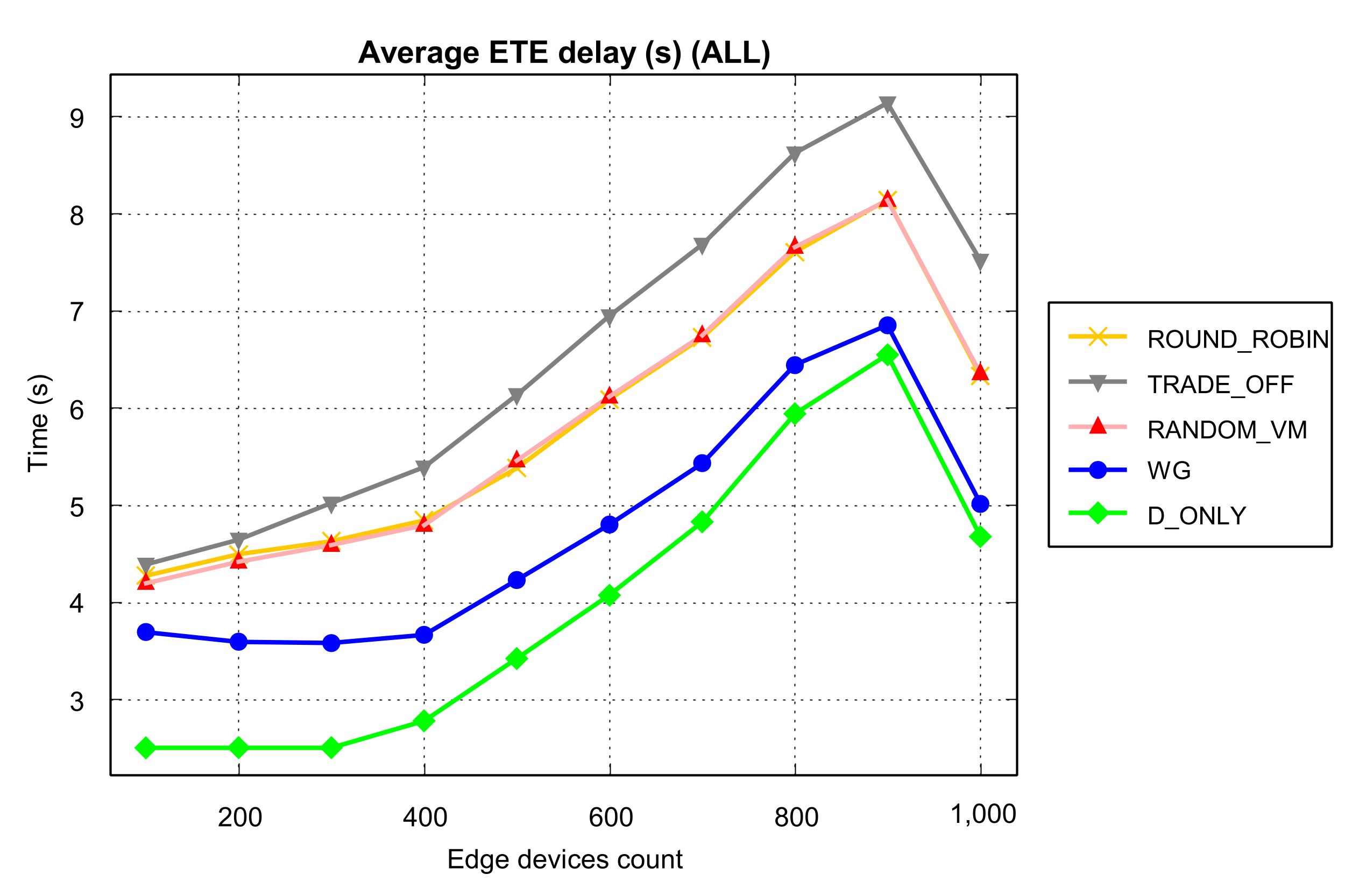}
\caption{Average End-to-End Delay of the orchestration algorithms}
\label{fig:s2}
\end{figure}

A thorough examination of Average End-to-End Delay across the spectrum of task orchestration algorithms, as shown in Fig. \ref{fig:s2}, revealed a common trend: an increase in end-to-end delay as the number of satellites expanded. Notably, our proposed Distance\_Only algorithm emerged as the frontrunner in this aspect, showcasing the best performance with an end-to-end delay variation ranging from 2.5 seconds to 5.5 seconds. In contrast, among the remaining algorithms, Trade\_Off exhibited the most challenging performance, with an end-to-end delay fluctuating between 4.5 seconds and 9.1 seconds.

\subsubsection{Satellites Energy Consumption}

\begin{figure}
\includegraphics[width=\linewidth]{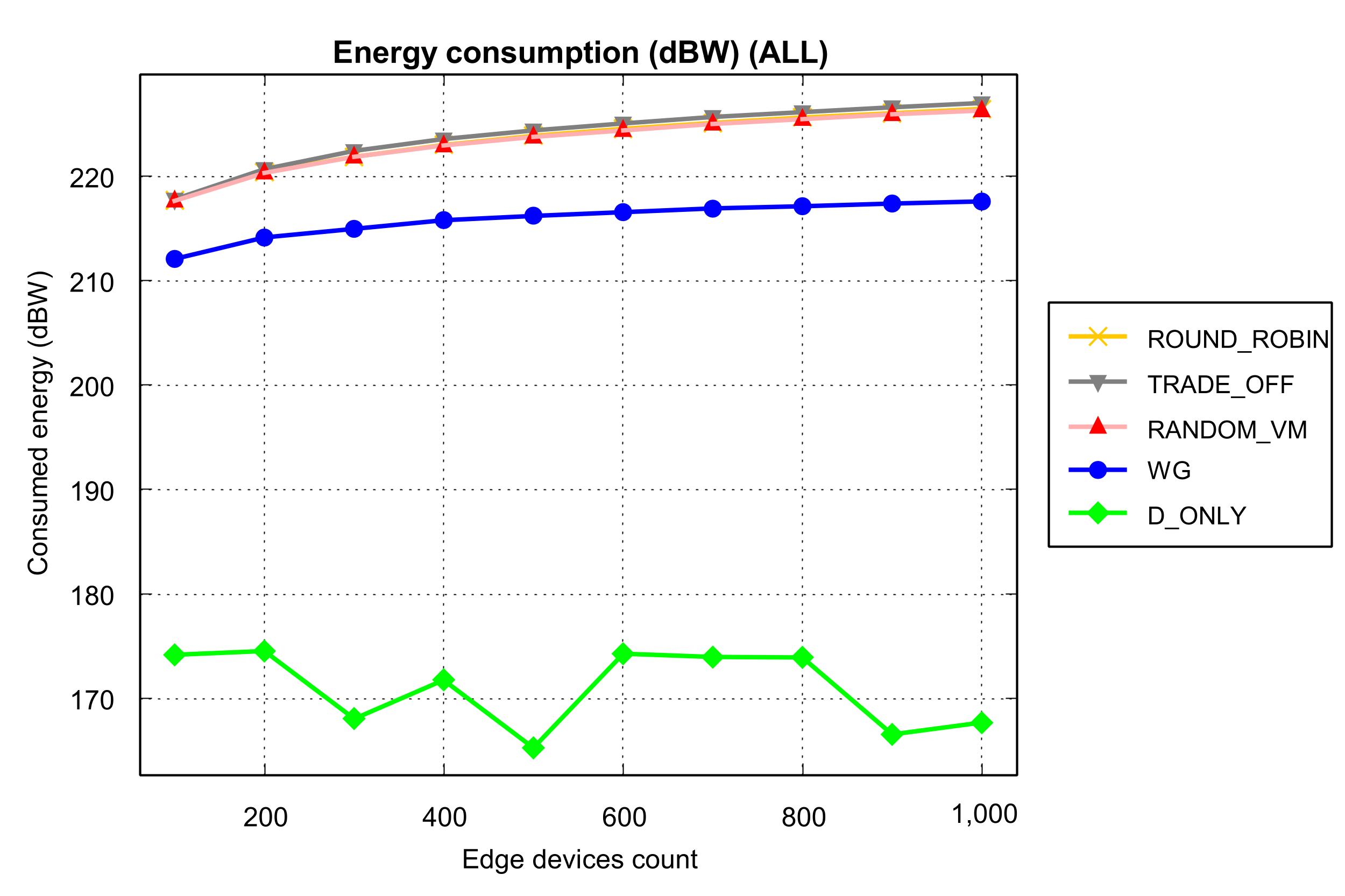}
\caption{Satellites Energy Consumption of the orchestration algorithms}
\label{fig:s3}
\end{figure}

The evaluation of energy consumption among the task orchestration algorithms that is shown in Fig. \ref{fig:s3} unveiled notable distinctions in performance. Our Distance\_Only scheme exhibited exceptional energy efficiency, maintaining a consistently low energy consumption ranging from 165 dBW to 174 dBW. This energy consumption exhibited minimal variation even as the number of satellites increased. In contrast, Round\_Robin, Trade\_Off and Random\_VM showcased notably higher energy consumption, with a marginal increase as the satellite count grew, resulting in values fluctuating between 218 dBW and approximately 237 dBW. The WG algorithm fell between these extremes, consuming energy in the range of 212 dBW to 218 dBW

\subsubsection{Tasks Success Rate}

\begin{figure}
\includegraphics[width=\linewidth]{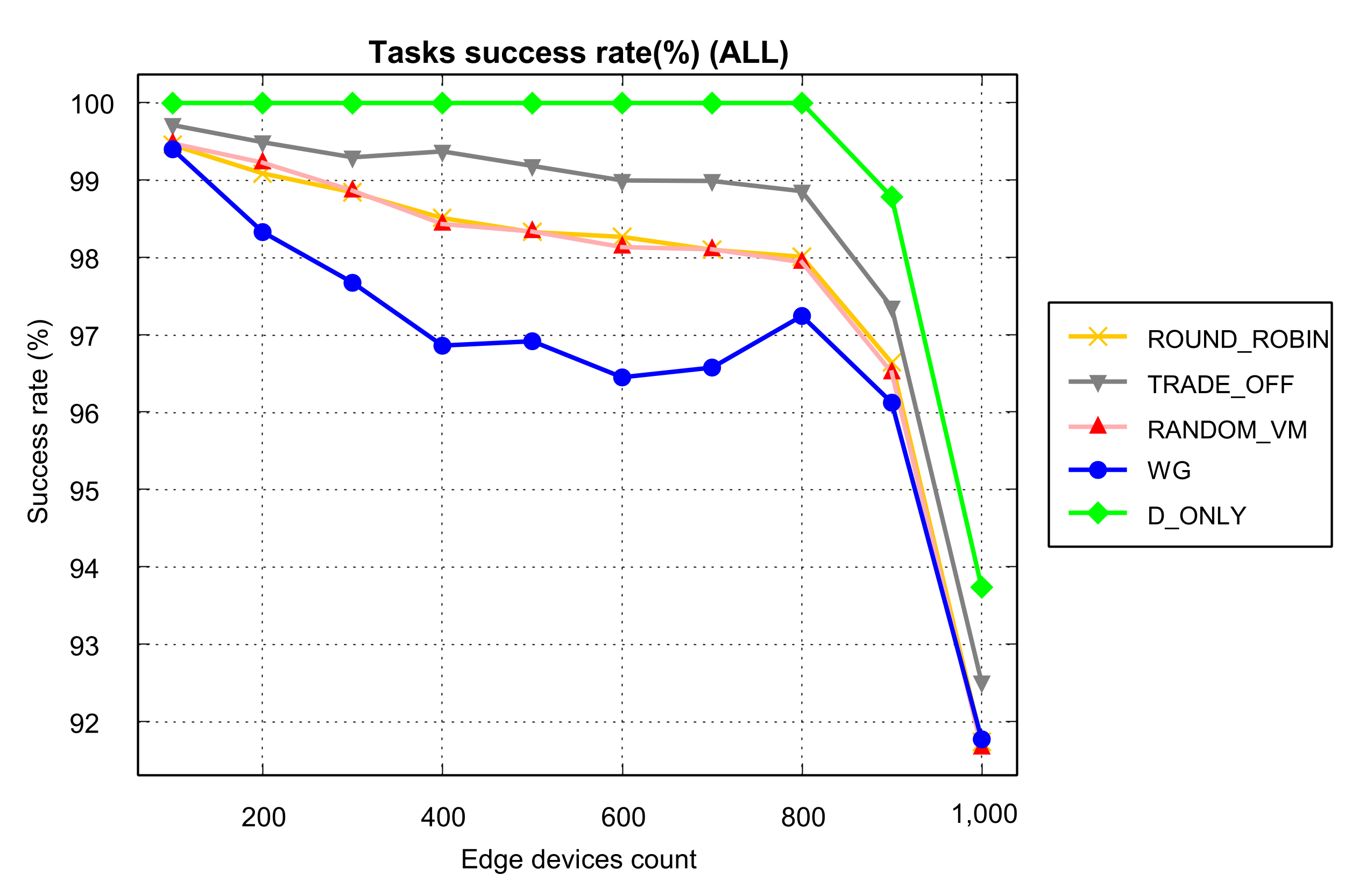}
\caption{Achieved tasks success rate of the orchestration algorithms}
\label{fig:s4}
\end{figure}

The examination of Tasks Success Rate across varying numbers of satellites, as shown in Fig. \ref{fig:s4}, highlighted intriguing trends among the task orchestration algorithms. Notably, our proposed Distance\_Only scheme showcased an impressive performance, consistently maintaining a 100 percent success rate while the satellite count increased up to 800 satellites. Beyond this point, a noticeable decline was observed, settling at 93.8 percent with 1000 satellites. This pattern was mirrored in Round\_Robin, Trade\_Off, Random\_VM and WG, commencing at approximately 99.5 percent and descending to below 93 percent as the satellite count increased. Additionally, it's worth noting that the WG algorithm exhibited the least favorable performance among the five schemes, characterized by a non-smooth variation between 100 and 1000 satellites.

\subsection{Discussion}

In this section, we provide interpretations of the results obtained from the four metric evaluations, shedding light on the performance of the task orchestration algorithms in a satellite-based mist computing environment.

For the VM CPU Usage, the relatively high CPU utilization observed in our Distance\_Only scheme can be attributed to its heavy reliance on mist-level satellites, which represent the lowest level in the architecture. By distributing heavy computing tasks among these satellites, the average CPU consumption naturally rises, surpassing that of other schemes that leverage resources from multiple satellite levels. While a high average CPU consumption may not be the most desirable outcome, it remains at a manageable level, which can be tolerated without causing critical issues.

In the Average End-to-End Delay, the Distance\_Only scheme excelled in achieving the best delay performance, owing to the inherent importance of satellite communication distances. Leveraging communication between satellites at the mist level ensures shorter delays, even when handling high-computing tasks. This efficient communication structure contributes significantly to minimizing end-to-end delays.

In the Satellites Energy Consumption, our scheme demonstrated commendable performance in energy consumption, achieving approximately a 20 percent of energy gain compared to other schemes. This can be attributed to the short distances between mist-level satellites, which are in close proximity to each other. This close proximity results in reduced energy consumption during satellite-to-satellite communication, contributing to more efficient energy usage.

Finally, in the Tasks Success Rate, our proposed Distance\_Only scheme exhibited exceptional performance by consistently maintaining a very high success rate, reaching 100 percent during the initial stages, in contrast to the other schemes. The significant difference can be attributed to the scheme's reliance on closely situated mist-level satellites, which minimizes task failures due to mobility, a common cause of task failure. However, the remarkable decrease observed as the number of satellites increased is due to execution delays of heavy computing tasks, which may not always be met, particularly with a higher number of satellites. This can lead to some mist satellites experiencing overloaded queues, resulting in delays that breach the task constraints.

At last, our Distance\_Only task orchestration scheme showcased a balanced performance, excelling in terms of end-to-end delay, energy consumption, and task success rates. While it exhibited higher CPU utilization, it remained within acceptable limits. These findings emphasize the efficacy of our proposed scheme in optimizing resource utilization and performance metrics in satellite-based mist computing environments.

\section{Conclusion and Future Work}
\label{s6}

In conclusion, the results of our comprehensive evaluation of task orchestration algorithms in a satellite-based mist computing environment underscore the significance of selecting the right approach to optimize resource utilization and performance. Our proposed Distance\_Only scheme, despite exhibiting higher CPU utilization, excelled in terms of end-to-end delay, energy consumption, and task success rates. These findings emphasize its efficacy in managing task orchestration efficiently.

As we look to the future, several directions for further research and development become apparent. One key area of exploration involves the development of adaptive variants of our Distance\_Only scheme. These adaptive approaches could account for the challenges arising from an increasing number of satellites, particularly addressing queue congestion and management issues. By dynamically adapting to the changes of loads, such variants can further enhance task orchestration performance.

Additionally, future research may delve into the utilization of cross-level satellites task offloading. This innovative approach would enable satellites, that use the Distance\_Only scheme, to intelligently balance their offloading by harnessing resources from other levels, including edge datacenters and cloud-level resources. By creating a dynamic, multi-tiered orchestration framework, we can optimize task allocation and resource utilization across the entire satellite-based ecosystem.

In the evolving landscape of satellite-based mist computing, our findings and future directions highlight the importance of continued research and innovation in task orchestration algorithms to meet the growing demands of modern satellite networks and ensure their seamless operation in a multitude of applications.

%%+++++++++++++++++++++++++++++++++++++++++++++++++++++++++++++++++

%+++++++++++++++ End

%\bibliographystyle{IEEEtran}
\bibliography{IEEEabrv,biblio}
%\biboptions{numbers,sort&compress}
%\bibliographystyle{elsarticle-num}
\bibliographystyle{unsrt}
%\bibliography{biblio}

%%\input{biblio.bbl}

\end{document}